# Science shops: A kaleidoscope of science-society collaborations in Europe

*Public Understanding of Science* (forthcoming)

Loet Leydesdorff [1] & Janelle Ward [2]

**Abstract**
The science-shop model was initiated in the Netherlands in the 1970s. During the 1980s, the model spread throughout Europe, but without much coordination. The crucial idea behind the science shops involves a working relationship between knowledge-producing institutions like universities and citizen groups that need answers to relevant questions. More recently, the European Commission has funded a number of projects for taking stock of the results of science shops. Twenty-one in-depth case studies by seven science shops across Europe enable us to draw some conclusions about the variety of experiences in terms of differences among disciplines, nations, and formats of the historical institutionalization. The functions of science shops in the mediation of normative concerns with analytical perspectives can further be specified.

[1] Université de Lausanne, School of Economics (HEC), 1015 Lausanne-Dorigny, Switzerland & University of Amsterdam, Amsterdam School of Communications Research (ASCoR), Kloveniersburgwal 48, 1012 CX Amsterdam, The Netherlands. Email: <loet@leydesdorff.net>; http://www.leydesdorff.net

[2] University of Amsterdam, Amsterdam School of Communications Research (ASCoR). Email: <j.r.ward@uva.nl>

# 1. Introduction

Danish undergraduate students test the quality of the water in ponds in a suburb because the inhabitants complain about the stink. The students provide suggestions to the local municipality which hitherto had neither the technological knowledge nor the resources needed to improve the situation. In Romania, an established scholar publishes about problems with the quality of surface water caused by a plant at the bend of a river. This project is organized in collaboration with colleagues from the Netherlands, and M.A. thesis students execute the work. A researcher at a public research institute in Valencia studies and advises about the incineration of meat remnants from industrial processes and the potential health risks involved.

These three settings seem to have little in common, except for the fact that they were all cited as examples of the best practices of science shops in twenty-one extensive case studies by an EU-funded project, INTERACTS.[1] Twenty-one in-depth case studies by seven science shops across Europe enabled us to comment on the variety of experiences. For example, what are the communalities and differences among these projects, and how can the results of such a wide range of projects be compared? How do the above-mentioned projects about health and safety issues relate to community-based research reported in the social sciences, or to a Master's thesis about poverty among children in Austria?

As a subcontractor of INTERACTS we had the opportunity to study these twenty-one case studies from the perspective of the research question of how the communication of science-shop collaboration can be improved. Can lessons be learned from science-shop practices about how to collaborate between clients with research questions and research capacities? We argue that the conditions of science shop practices have been changed because of the ICT revolution and particularly the widespread availability of scientific information on the Internet. While previously these shops had mainly aimed to maintain a local window institutionally, the possible communication of results to wider audiences via the Internet makes them increasingly a potential instrument in the process of the public understanding of science (European Commission, 2001). The shaping of public demand for science and technology may contribute to strengthening a knowledge-based economy by opening windows for innovative action (Leydesdorff & Etzkowitz, 2003; Cooke & Leydesdorff, 2005).

From this perspective of science communication, the cases were evaluated in order to determine the visibility of the science shops, specifically in terms of the visibility of the results, the NGOs, and the researchers involved, both on the Internet and in on-line databases like the *(Social) Science Citation Index*. The current study expands on some of these issues. However, before presenting the current work, a brief description into the reality of science shops is called for.

---

[1] INTERACT stands for "Improving Interaction between NGOs, universities and science shops: Experiences and Expectations." The project involved seven science shops in Europe, each of which analyzed three cases of "best practice" in depth by interviewing relevant NGOs, students, and researchers, as well as mediators and university administrators, in order to generate ideas about how to improve these practices. The full reports are available from the website of the project at http://members.chello.at/wilawien/interacts/reports.html. INTERACTS follows up on a previous EU project called SciPas that produced a number of policy reports for the European Commission in the period 1999-2000. The reports of SciPas are available at http://www.livingknowledge.org.

*Science shops: ideas and recent developments*
The science-shop model was initiated in the Netherlands in the 1970s, and spread throughout Europe during the 1980s, but without much coordination. The crucial idea behind the science shops involves a working relationship between knowledge-producing institutions, such as universities, and citizen groups that need answers to relevant questions. In offering this ideal, bottom-up approach to research, the hope is that the relationship between science and the public can be encouraged through accomplishing active collaboration in research, as well as providing such groups with access to the university and its facilities, regardless of institutional barriers (Bunders & Leydesdorff, 1987). Even beyond this mediation between science and society, the science-shop model also wishes to empower the clients with the insights provided.

      The diversity and scope of questions is such that the most successful centers are having difficulty in satisfying demand. The science shops would gain from cooperation, with the aid of the Commission, in pooling their resources, their work, and their experience. To this end, the European Commission has funded a number of projects for taking stock of the results of science shops.[2] Because of their combination of local and European elements, the Commission placed the science shops on its agenda in the framework of its *Science and Society Action Plan* (European Commission, 2001), and encouraged them to collaborate with each other:

> There are in Europe various types of science shops close to the citizen in which science is placed at the service of local communities and non-profit-making associations. Hosted by universities or independent, their common feature is that they answer questions from the public, citizens' associations or NGOs on a wide variety of scientific issues. The first science shops were opened in the Netherlands in the 1970s and the idea was then taken up by about 10 other countries throughout the world. There are now over 60 science shops in Europe, mainly in the Netherlands, Germany, Austria, the United Kingdom and France. (*Ibid.*, at p. 15)

The Commission provides financial support and also stimulates the development of a network of science shops with a website and other publications, both virtual and in hardcopy.

*Present issues: Collaboration, comparative analysis, science-shop culture*
Here, we intend to go beyond the preliminary conclusions reached in the INTERACTS project by expanding on three stimulating themes. First, we examine the model of collaboration proposed by the science shops. How has this model evolved over time, and how can the process itself further be improved? Next, we will focus on comparative aspects of the case studies presented in the INTERACTS project. What specific similarities and differences exist? Finally, we provide an interpretation of the culture that encompasses the science-shop community. This study focuses on what one can learn from these case studies for the "new social contract between science and society" that has been advocated by scholars in the tradition of science and technology studies (Caracostas & Muldur, 1998; Etzkowitz & Leydesdorff, 2000; Gibbons, Limoges, Nowotny, Schwartman, & Scott, 1994; Nowotny, Scott, & Gibbons, 2001).

---

[2] The European Commission has decided to support a new network of science shops called ISSNet ("International Science Shops Network"). The website of the EU about science shops is located at http://europa.eu.int/comm/research/science-society/scientific-awareness/shops_en.html. A color brochure can be downloaded at http://europa.eu.int/comm/research/science-society/pdf/science_shop_en.pdf

## 2. Case studies: background information

Seven science shops across Europe provided us with 21 case studies involving their research. Here, a brief summary of some relevant aspects of each of these science shops, country by country, is in order.  Each country provided case studies from two to three science shops.

*Denmark*
The Danish case studies involved engineering students at the science shop at DTU (The Technical University of Denmark), as well as another science shop at RUC (Roskilde University Centre). Both science shops are related to the central administrations of their respective universities with the purpose of providing inroads into both science and research for organizations in civil society, and, *vice versa*, to grant students the possibility of including "real life" topics as part of their curricula through cooperation with these groups (Zaal & Leydesdorff, 1987).
	RUC provided one study that involved a literature review and interviews with cyclists, politicians, and traffic planners, in order to understand their perception of the bicycle as a technology. The project involved two M.Sc. engineering students in their fourth year. The Danish Cyclist Federation wished to improve its position in the discussion on traffic planning by understanding the social construction of the image of bicycling.
	The second case study involving DTU related to a request for assistance with research regarding environmental management in a day-care center. The research was carried out by two students midway through their engineering studies, and was accomplished through a literature review and through informal talks with the staff at this day-care centre.  Finally, the third case drawn from RUC, engaged theoretical considerations regarding bio-manipulation in shallow eutrophic lakes, as well as tests and water samples taken from the village pond of Litte Rørbaek. Four Environmental Biology students, with high-quality supervision in their fourth semester, carried out the project.

*Austria: Tyrol*
Three science shops were involved in the case studies for Innsbruck.  One, the Institute for "Forschung, Bildung und Information" (FBI), is a non-university based science shop and research institution.  It serves as a link between academia and society as well as between theory and practice, on issues related to research, society, and culture, with a special focus on women and gender issues.  The second, the Wissenschaftsagentur Salzburg, a university-based science shop located in Salzburg, is organized as a non-profit organization; and the third, Patenschaftsmodell Innsbruck (PINN) is a science-shop equivalent, or more specifically a service center for enterprises and organizations at the level of the faculty.
	The case study from the science shop PINN evaluated customer satisfaction in relation to a service called "Mediation in penal matters," which was provided by the NGO. Two final-year undergraduate students in the Faculty of Social and Economic Sciences conducted the evaluation within the context of their Master's thesis, and under the supervision of academic staff.  The second project, implemented through the Wissenschaftsagentur Salzburg science shop, was intended to provide a well-grounded scientific basis for the stimulation of youth work, for the establishment of a youth center and youth information point, and furthermore for the initiation of future youth projects in this region. One final year undergraduate student conducted the study. The third case study resulted from an appeal by an NGO located in the health sector with a focus on women, to the science shop, the Institute FBI. The project evaluated a series of lectures involving precautions against heart disease for Turkish migrant women in Tyrol. The evaluation was conducted by two researchers who were staff members at the Institute FBI, as well as two medical students of Turkish origin, who worked as interpreters and experts in the cultural background.

*Austria: Vienna/Graz*
Two cases were chosen from the science shop Graz and one from the science shop Vienna. All three cases represent the cooperation of NGOs with universities, researchers, and students. One of the cases concerned a larger project with several collaborating student researchers, and the other two included one or two M.A. students.

The first case involved an NGO and the science shop Vienna, and focused on empowerment in a neighborhood. The aim was to improve the lives of tenants in this settlement by giving them more responsibility and autonomy. The project is part of a series developed by an active researcher of this institute, and a Master's thesis was central to the case study report. The second case study dealt with the importance of finding volunteers to provide companionship and to improve the quality of life for mentally disabled persons. Specifically, this NGO wanted to know if such social companionship is successful. The project's intermediary was the science shop Graz, and the student completed a Master's thesis with professional supervision. Science shop Graz presented also the third case study, which dealt with an NGO interested in receiving sound scientific information on government subsidies for families, with a focus on improving the conditions for children in these families. From an academic perspective, this project was very successful: the department hired one of the students, and the other has co-authored a publication with the supervisor.

*Germany*
The German case studies represent two science shops: *kubus*, the Kooperations und Beratungsstelle für Umweltfragen (Co-operation and Consulting Office for Environmental Questions), a science shop located at the Technical University Berlin, and the science shop Bonn (Wissenschftsladen Bonn), which focuses on ecology and environmental protection.

The first case study dealt with the construction of the Tiergarten Tunnel in Berlin. Kubus was contacted for the formulation of an expert report, or environmental impact study, in an NGO's legal case against the tunnel project. Two students of this research group received an award from the "Love Parade" for an M.A. thesis about the damage resulting from these construction activities in the Tiergarten. When the Tiergarten Tunnel project led to conflicts among the NGOs, the idea for the KREKO project was formed, resulting in the second case study. This project addressed problems within and between NGOs, environmental groups, and environmental associations, and also worked to improve internal communication and cooperation. The third case study's goal was to create modules for a Germany-wide information, cooperation, and development network dealing with foundations in the field "Environment and Local Agenda 21." The science shop in Bonn was the organizer of these collaborations.

*Spain*
Three different science shops made up the case studies in Spain. The first, Pax Mediterranea, deals with research within the social arena of ecology, economic development, and social cohesion strategy, with environmental and socially sustainable perspectives. It is actively involved in various European and local research and monitoring projects and observatories. The second, Architecture and Social Commitment (ASC), is concerned with social instruction in the universities, the construction of a sustainable habitat in inner cities, equality on a global scale and the instruction of citizens, without a primary reliance on trained architects. Finally, science shop ISTAS (Istituto Sindical de Trabajo, Ambiente y Salud) is a self-funded technical foundation promoted by the Spanish Trade Unions Confederation (CC.OO). It supports social activities to improve working conditions and environmental protection in Spain.

The first case study originated from a group of "green" organizations who contacted Pax Mediterranea hoping to carry out an independent study of the ecological issues that were present in Seville society. These groups organized in order to look at possible future scenarios where their input and action could be called for (Roja, 2001). The second project began when the NGO Human Rights of Andalusia were informed that a Romany (Gypsy) shanty neighborhood was being removed from public/private land in order to make room for building contractors; the NGO contacted the science shop ASC. The third case study dealt with the issue of burning cattle meat in kilns, as this had become a risk of concern both to workers and for the environment. ISTAS made a recommendation to CC.OO to study the issue, for the sake of minimizing environmental risk to workers.

*United Kingdom*
The UK case studies involved two science shops: Interchange and Student Link. Interchange involved students from three local universities—Liverpool University, Liverpool Hope University College, and Liverpool John Moores University—all undertaking research projects with local NGOs. Student Link at the University of Wolverhampton enables senior undergraduate students to conduct applied research projects for one or two semesters.

The first case study consisted of four health-related projects involving Benington Hospital, all utilizing graduate students. The main aim of the second project was to provide an independent evaluation of a day center for older people, from the service-users' perspective. The project began with a request from the NGO for an external evaluation of the Day Centre. Two undergraduate students chose this subject for their applied social research project. The third case study was part of an ongoing relationship with Age Concern. The participants in the project consisted of one senior undergraduate student at the University of Wolverhampton, the academic supervisor, the coordinator of Student Link, the science shop, the Befriending Service coordinator, and the manager of the NGO. The report has been used as a guide for incoming coordinators.

*Romania*
There are two science shops involved in the Romanian case studies. InterMEDIU Information Consultancy and the Department of Environmental Engineering at the Technical University of Iasi were responsible for two of the case studies. This InterMEDIU is based in the Faculty of Industrial Chemistry, and was funded as a result of a bilateral cooperation agreement with the University of Groningen in the Netherlands. The second science shop is InterMEDIU

Information and Research Centre, "Al.l.Cuza," at the Faculty of Biology, at the University of Iasi.

The first case study was a pilot project of the science shop InterMEDIU (Technical University of Iasi).  The aims of the project included consultations with the community about drinking water, a comparison of the qualitative problems raised about the situation in the treatment plants, the formulation of proposals to improve the existing situation, and the organization of a public debate concerning the quality of drinking water.  Students in the Environmental Engineering Department were given an opportunity to apply their knowledge relating to Water Treatment technologies, and also to learn more about the techniques of social inquiry, project management, and computer applications.  The second case study began with a request from an environmental NGO, the Ecology and Tourism Club Moldavia.  The project's objective was the evaluation of the environmental impact of the waste waters generated from yeast production on the receiving waters of the river Siret.  The M.A. thesis involved was awarded a prize from the faculty. The NGO used the information presented in the report both for the members of the NGOs and to provide information to the local community.  The third case study was requested by three NGOs, and involved the InterMEDIU science shop and university staff and students from the Faculty of Biology. The research for this project continued previous studies carried out between 1995 and 1998 by the Romanian Ornithological Society, which involved a full biological documentation for a RAMSAR site assessment in the area of the Wetland Vladeni.

*Case studies: observations*

These brief descriptions of the case studies analyzed in the INTERACTS projects raise several questions of relevance to the work at hand.  One issue relates to the idea of intermediating between clients and researchers and/or students.  How do these science shops define their roles?  How do the structures of the institutional organizations involved affect the model?  What about the missions of these organizations (for example, higher education versus academic research)?  Many of the projects at hand are a result of students' work, and the final reports are often presented only in these students' M.A. theses.  Is the possibility of increasing social capital in these science shops harmed by this mainly "gray" literature representing the outcomes of the projects?  These questions will be addressed in the remaining sections of this paper.

**3. Science shops as mediating agents**

Since the emergence of the science shops in the 1970s, the model has been diffused from the Netherlands within Europe and even beyond (Farkas, 2002; Fischer & Wallentin, 2002; Irwin, 1995; Mulder, Auf der Heyde, Goffer, & Teodesiu, 2001). From its very beginning, the Dutch model was based on coalitions among various groups (Wachelder, 2003). For example, some of the Dutch science shops (e.g., Utrecht) were heavily engaged in political actions outside the university context, while others (e.g., Amsterdam) considered the shops primarily as an option for institutional reform within the university context. The University of Amsterdam experimented with developing science-shop questions into longer-term university research policies, while the group in Utrecht wished to remain on the oppositional side (Leydesdorff, Ulenbelt, & Teulings, 1984).[3]

The variations among disciplines, political contexts, and institutional formats require the science shops to make pragmatic trade-offs. The crucial inquiry in the intermediation is how the capacity of public research can be used to solve social and environmental problems, and conversely, how stakeholders in these problems can provide access to questions and domains for the various layers of the university and its respective missions of higher education and scientific research. However, the distinction between being enrolled institutionally or remaining outside of the university setting can also be observed in the different positions of the current science shops: do they wish to operate within a given university context, or do they try to change the academic arrangements? Answers to these questions can be expected to vary between more established disciplines (e.g., chemistry) and more action-oriented fields such as women's studies.

The science shop in Innsbruck (the FBI Institute), for example, profiles itself as a knowledge base and an independent center of expertise on the side of action groups and NGOs, but science shops elsewhere in Austria mainly provide university students with possible topics for their Master's theses. The outputs resulting from these varying configurations can be different because the institutional roles and the professional expectations differ at the structural level. In a report about science shops in Europe, the INTERACTS project consortium argued that the mediating role is itself under pressure to change because the relations between the two sides of the mediation are experiencing modifications over time. Four waves of science shop work that were distinguished in the state-of-the-art review of INTERACTS (Fischer & Wallentin, 2002; Fischer, Leydesdorff, & Schophaus, 2004). (Later, these waves will be discussed in a more historical and cultural context.)

1. The initial wave consisted of mainly of Dutch science shops during the 1970s. These developments commenced from within the university system, formed by coalitions of progressive staff members together with activists in student movements. The terms of the debate were set by the science policy discourse about "democratization," that is, access to higher education and university research as the scientific knowledge bases of society.
2. A second wave (in the 1980s) was strongly interwoven with the further institutionalization of alternative movements like the "Bürgerinitiative" in Germany. These groups in civil society (i.e., outside academia) needed to develop their knowledge base and sometimes turned to the university for assistance. Furthermore, some of these NGOs recruited membership among students and university staff members. Thus, a common perspective on the science shop practice was developed.
3. A third wave during the 1990s was based on the increased awareness of the need to build social capital and to fight exclusion mechanisms in post-Cold War societies. From this perspective, the network function across institutional boundaries became an objective in itself. Science shops provided a model for engaging the university in non-economic objectives with groups that were hitherto excluded from these knowledge-intensive domains. While the first wave was mainly grounded in the critique of students and staff at the universities and the second was mainly positioned in civil society, during the third wave this gap was bridged on the basis of professional considerations. Therefore, this wave was mainly driven by social scientists. One example of this latter type of project is community-based research as practiced by science shops in the U.K. (Hall & Hall, 1996, 2004). Through the disciplinary dimension of community-based research, relations with similar groups in the U.S.A. and Canada functioned as an important resource for the movement during the 1990s (Sclove, 1995).
4. A fourth wave existed in parallel with the third, but can be distinguished because the developments took different shapes in the countries of Eastern and Middle Europe. Perhaps we can also place the science shops in South Africa and Third World countries in this context (Mulder *et al.*, 2001). Groups and organizations in these countries were able to take advantage of this model of science shops during the reconstruction of their economies and their societies. In such situations one may locally be able to explore the synergetic potentials of new arrangements. In Spain—one of the case studies of the INTERACTS project—one would expect to find an early example of this construction of civil society after the Franco period. The knowledge base of society can be used as a cultural resource for the reconstruction of institutional arrangements.

---

[3] Although most science shops in the Netherlands have been marginalized (Wachelder, 2003), the Catholic University of Brabant in Tilburg still develops science-shop questions into Ph.D. projects to be executed with support of the university.

Bridging mechanisms like science shops at the institutional level can be useful in developing social integration. Functionally, these intermediaries may also provide a counterbalance and legitimacy in a context where more commercially oriented technology transfer and science parks are supported for economic reasons. One can expect tension between the public and the private domains over matters of access and principles of appropriation (Leydesdorff & Etzkowitz, 2003). Therefore, one would suppose the mechanisms of transfer also to be different.

    During the third and fourth waves that took place in the 1990s, professionalization of the transfer of knowledge became crucial because the creation of social capital is based on the generation of trust. In this context, the quality control of projects could no longer be left to undergraduates and student volunteers at lower levels of the organization. The university itself had to make a visible commitment to the objective of generating social capital. All science shops under study took quality control as a very serious issue, for example, in terms of the demands for supervision of the M.A. students involved. However, the building of social capital cannot be addressed properly at the level of a single science shop as a source of the information without access to relevant diffusion mechanisms.

    The introduction and spread of the Internet during the past ten years has noticeably changed the arrangements for the diffusion of scientific knowledge. The science system itself is deeply affected by the ICT revolution, as communication is central to scientific work (Gibbons *et al.*, 1994; Leydesdorff, 1993, 2001; Luhmann, 1990). The external relations have changed because increasingly, stakeholders have direct access to the relevant knowledge bases. This new configuration is unlike previous arrangements as the costs involved in gathering information have declined. As a result, the configuration has altered in which science shops can add surplus value to the relevant information or knowledge.

    In the meantime, the science shops have evolved in terms of the forms that are sustainable in their present environments, and furthermore in terms of the specific forms of integration that have been functional during the last decades. It is therefore not surprising that there is a flux of new entrances and exits among science shops. A reflection of the current state

may accordingly provide us with a variety of elements that can be crucial for "a new social contract between science and society" from a perspective not exclusively economic. Science shop practices offer a point of crystallization where the latent demand for public access to science and technology can become visible, not only as an intellectual interest (as in a science museum), but as an interest rooted in social structures, e.g., NGOs. The clients of the science shops formulate a need to obtain access to research capacity, but they provide a kaleidoscope of issues. Can formats be recognized that inform us about the structures of these demands?

*A new social contract of science and society?*
The science-shop case studies teach us that a new social contract between science and society will have to encompass a multitude of intellectual shapes and institutional formats. This process is not one-sided, as with a supply-driven client relation (such as technology transfer), because the knowledge base is expected also to structure the configuration for both sides. Knowledge-based insights and innovations are not like commodities freely available on an anonymous market, but can be considered rather as highly specific and codified channels of communication between science and the public. For example, a client may turn to a university or science shop with a specific question for which no answer is yet available. The question may, however, be very relevant from the analytically different perspective of a specific research program.

How this mediating role can be institutionalized is another issue: the institutional framing requires a further reflection on the local contexts. At the institutional level one can expect specificity in the windows of communication to be further developed. For example, within the higher education system teachers have an institutional need to provide students with relevant and interesting research questions for their Master's theses. This point of entrance for client topics can easily be recognized as a place where the higher-education system can provide for a specific form of delivery of expertise to social groups. Unlike Ph.D. projects, theses at the Master's level are usually not expensive (because of, for example, internship relations). Furthermore, if sufficiently supervised, the results can be used as sources of legitimacy in political discourse, while in technology transfer economic considerations require a higher degree of reliability and planning. The time pressures are also of a different kind.

But this is not the only perspective. Conversely, one of the Spanish cases shows the need of trade unions to build up in-house expertise that is backed by the research system, but not necessarily at the university level. Here, their extra-university expertise calls upon a public research institute to exercise a quality-control function for the relevant knowledge. In a knowledge-based society, the organization of scientific knowledge is needed in domains other than academia. In such cases, whether organizations and groups in the civil society will be able to turn to academia with their questions depends on the development of relations of mutual trust. Although the science shops provide a low threshold, NGOs that have developed an internal knowledge base may prefer to discuss and negotiate with scientific departments and institutions without mediation. This is particularly likely if the NGO involved has in-house academics that already belong to their own intellectual networks.

## 4. Comparative aspects of the case studies

Given the local character of science-shop activities, it can be expected that, after a few decades in operation, science shops have fashioned their own "best practices." These best practices vary noticeably given the differences in their relevant environments. Science shops can often be identified by the interests of one of the parties involved: the clients, the social justifiability and the embeddedness of university research, or the concern for students in higher education.

*National and disciplinary differences*
National differences among science-shop practices and the disciplinary affiliations of the researchers provide two major dimensions for comparison among the cases. Perhaps, with the exception of Spain where the "science shop" is not yet itself a concept used to assist the cooperation, the common origin of the discourse about science shops in the various European countries is recognizable. These activities seem to attract highly motivated, culturally advanced, and socially engaged students and young scholars who are seeking to pursue careers that are intellectually and socially meaningful. As discussed in the previous section, the science shops provide and generate social capital in terms of relevant networks, first of all for the researchers involved. These projects can perhaps be considered as a distributed format of the new social contract between the universities involved and their environments.

The mediating role of science shops is institutionally recognized to such an extent that often the science shops can, for example, turn routinely to other offices within the departments that help students find internships. These provisions are standardized in departments in the social sciences more than in the natural sciences. Some of the shops are closely related to specific departments in applied natural sciences. Thus, the disciplinary categorization can become more important than the national one for the mechanisms that prevail in the collaboration.

For example, while differently structured in terms of their institutional organization, the science shops of Vienna and Liverpool provide in some respects similar services to both M.A. students and clients. While the team in Liverpool is highly focused on its disciplinary orientation in the social sciences, the wider range of supervisors in the Vienna cases are intellectually organized in similar frameworks. The two groups may not know each other's work because most of the publications by the Austrian scholars are in German, but their orientations in terms of "grounded theory" (Glaser & Strauss, 1967), appreciation of non-economic objectives, supervision of students, etc., are highly comparable. Thus, while the institutional settings are very different, the intellectual orientations and relevant theorizing operate along similar lines.

A second group of science shops is focused on "environmental issues," but here a further distinction can be made. The science-shop activities in Romania, Denmark, and probably a number of the Dutch shops (which were not included among the cases studied in this project) can be subsumed under the heading of "environmental engineering and management" (Teodosiu & Caliman, 2002). The structural position of these problems in Western and Eastern Europe is different, but this field is still an important driver of science shop activities in both research and higher education.

Although environmental issues have been a driver in organizing science shops in Germany since the 1980s, the German case studies, like the Spanish ones, are less oriented towards higher education and more towards the professionalization of consultancy by members of the scientific staff. In these two countries, science-shop questions have offered access to intellectual domains that were experienced as challenging by individual scholars. In both the German and the Spanish cases we found a Ph.D. thesis among the case materials in addition to international publications (Barsig, 2002; Roja, 2001). The professional activities in these two systems seem institutionalized. The Innsbruck institution FBI is careful to prevent further institutionalization, but it can still be compared with a professional unit in these terms because of its sustained focus on scholarly publications (e.g., Schweighofer-Brauer, Schroffenegger, Gnaiger, & Fleischer, 2002).

In summary, three substantive foci can be distinguished among the twenty-one case studies:

1. "Environmental engineering and management" both as a disciplinary profession and in terms of higher education (Teodosiu & Caliman, 2002);
2. Social work, its problems, and its evaluation aimed at the improvement of intervention strategies (Hall & Hall, 1996);
3. Professional expertise in the case of problems that cannot easily be disciplined, such as the meaning of cycling or the ecological and social effects of building a tunnel under the Tiergarten.

These intellectual foci interact with two types of structural relations that can be maintained from the side of the university, one based on the higher-education function of this institution and the other based mainly on its research function. The questions from clients of science shops provide the higher-education system with topics, for example, for writing Master's theses. The university could consider using these topics and theses as starting points for the institutionalization of new research lines, but this is currently not a standard practice. In most cases, the individual students are successful in using the advantages of their specific projects in furthering their career either within or outside a university setting.

The topics can be developed into research questions if the interface between the clients and the science shops has been further professionalized. This professionalization can occur on either side of the interface. In already professionalized configurations, the topics are provided with a specific interpretation in terms of a scientific specialty before the research process has begun. If the professionalization is developed on the side of the client, then specific networks are used for the mobilization of expertise. But if the professionalization has mainly taken form on the side of the science shop, the latter tends to function as a consultancy operation with a low threshold for economically weak but politically urgent demands.

In the Romanian cases we found a more integrated approach in terms of the higher education and research functions of the university. This may relate to the specific phase of the development of civil society and the discipline of environmental engineering in Romania. In this country, networks seem to be able to mobilize across the institutional divides of established institutions. The system can perhaps be considered as less differentiated and fragmented than in some of the Western European countries at this stage. Perhaps this emerging form of organization can be developed into a more stable advantage for the transitional economies.

If the university would like to profit from societal input both at the level of higher education and at the level of research, communalities in the interfaces of research and higher education with the university environment should be further developed. A few points will later be elaborated among the recommendations, such as the establishment of rewards for best practices. Such evaluations may have to distinguish between social relevance and scientific quality. A university may wish to establish a standing committee at the level of the board that investigates the potentials for further development at the interface with the surrounding society more systematically in terms of research and higher education. The European Commission could provide universities that wish to move in this direction with start-up subsidies (on a competitive basis) in order to strengthen the implementation of its own *Science & Society Action Plan*.

## 5. The cultural contexts of science shops

From a historical perspective, it is important to examine how the four waves of institutionalization of science shops can be considered as results of changes in the relations between science and society. These changes at the regime level affect the conditions of the institutionalization (Giddens, 1984). Let us return to the four waves which were distinguished by the INTERACTS project (Fischer & Wallentin, 2002).

The first wave and the student movement can be recognized, with hindsight, as a product of the welfare state that emerged in Europe in the 1960s. The student movement was culturally motivated to use science for purposes other than economic advancement. This was often expressed with the slogan of the 1968 revolt, "*l'imagination au pouvoir*" ("power to the imagination"). The experiments of science shops in the 1970s can be compared with other attempts to generate bridges between a left-wing (socialist) appreciation of science and technology, and further reflection among intellectuals about cultural changes in the role of scientific knowledge in society. These roles were changing because knowledge production was becoming increasingly organized and controlled (Habermas, 1968; Whitley, 1984); national science and technology policies in this period were still very much under construction.

For example, in the 1970s the Labour Minister of Industry in the U.K., Tony Benn, endorsed the so-called Industrial Workers Plans (Cooley, 1980). The German Ministry of Science and Technology (BMFT) launched an ambitious research program for the "Humanization of Labour," and alternative product designs involving users at an early stage became almost a benchmark of Scandinavian quality. The French "colloque national" on science and technology in 1981 can also be considered as part of this development (Vavakova, 2000; Leydesdorff & Van den Besselaar, 1987a, 1987b). Similar discussions concerning the role of scientifically organized knowledge in the economies of Eastern Europe took place in Czechoslovakia during the "spring of 1968" (Richta *et al*., 1968)

Perhaps, one can identify this model as the "old left" (marxist or neo-marxist) as against the "new left" or "green" models of the 1980s. The first wave focused on institutional reform, and the second model also had an anti-institutional flavor. The environmental movement considered science and technology as themselves part of the problems caused by industrialization, thus creating a paradox for science-centered institutions to be trusted to provide solutions to the problems formulated by the environmental movements. In practice, however, the two models have continuously been recombined, but for the sake of analytical clarity it can be useful to highlight these distinctions.

The second model emerged in relation to the decline of the industrial model for economic development. In the advanced economies, the emphasis had shifted from science policy to technology and innovation policies (OECD, 1980). While low-wages countries were able to out-compete the OECD countries in terms of labor and raw materials, the advanced industrial systems increasingly turned to knowledge-based innovations to maintain their competitive edge. During the early 1980s, this evolution led to a rethinking of industrial policies under a neo-liberal regime that was strategically oriented towards "re-industrialization" (Rothwell & Zegveld, 1981). The environmental movement was thus pushed into the position of having to defend the environmental legislation of the 1970s.

Although the alternative movements remained ambivalent about scientific rationalization, some elements of the environmental movement found shelter in the university, and it become a breeding place for cultural reflection and for the development of alternative models (Beck, 1992). University departments of environmental sciences had been institutionalized successfully during the 1970s, but in many institutes these "interdisciplinary" units remained vulnerable to budget cuts given the new regime that focused, for example, on the development of entrepreneurial activities like "biotechnology" and ICT.

Our case materials provide more examples from this second wave than from the initial model, but sometimes the cases can be analyzed in terms of a combination of all three movements: the alternative movement of the 1980s with its strong focus on environmental issues, the cultural reform demanded by the student movement (which is still very much alive in Berlin), and the trust in science and technology as potential forces of transformation and enlightenment viewed from a more traditional perspective. As noted previously, a fourth element has to be added because of the more recent focus on the use of bridging institutions for the generation of social capital. In this regard, a disciplinary perspective within the social sciences, notably a focus on "grounded theorizing," found an institutional form in activities that are very akin to the science-shop model (Hall & Hall, 2004).

It is now possible to appreciate these different dynamics as cultural (sub)dynamics that can continuously be recombined in the complex evolution of social relations (Luhmann, 1984). The case studies illustrate how these different resources and repertoires are parts and parcels of European culture. For example, the "old" model of anti-capitalist opposition is still important in some of the Spanish case studies. In the Austrian case studies the concern for groups marginalized by capitalist development can also be considered as a driver, but now with a focus on "inclusion." All the cases can be considered as examples of the construction of social capital.

The emphasis on the value of networking itself has been typical for the 1990s. Fukuyama (1999), for example, discussed the crisis of modern society in terms of the erosion of social capital (Putman, 2000). From this perspective, network formation can be considered as an objective in itself because networks provide a mechanism for preventing social exclusion. Such social networking, however, cannot be achieved by deliberate policies at the national level, but has to be made possible within civil society, for example, in terms of new interactions among existing institutions.

*Shaping public systems of innovation*
The gradual decrease of identity at the national level was reinforced in Western Europe by the end of the Cold War, the subsequent disappearance of the Soviet Union and the reunification of Germany, and—last but not least—the emergence of the Internet. The concept of the "new economy" became virtually synonymous with a knowledge-based economy that is innovation-driven and globalized. While globalization is sometimes perceived as a threat, innovation can be also turned into a celebration of community formation because the new products and processes have to be accepted and disseminated locally. The shaping of new conditions enables people to cross bridges across ethnic groupings and over institutional (and national) divides. This spirit of new options, a result of changing competitive conditions across borders, can be considered central to the EU project itself.

From the perspective of prioritizing the generation of social capital, the role of the collaboration changes again. The networks generate knowledge endogenously in terms of consensus and/or non-consensus formation, exchanges of convictions, rational expectations, and arguments. High-quality communication across borders becomes more important than providing expertise and counter-expertise in an oppositional mode. The university can be expected to play a constructive role in these exchanges, while the partners can accept that the university has to cultivate roles like guarding the quality of scientific communication at the global level and providing society with qualified personnel through higher education. Given these global objectives of the university, the institutional parameters of the operation can be debated and reconstructed among the partners in terms of locally optimal conditions.

For example, some universities have been developed with venture capital during the 1990s into "entrepreneurial universities" (e.g., Chalmers University in Gothenburg; cf. Clark, 1998), while others—for example, those located in less favorable conditions—have attempted to act as regional innovation organizers (Etzkowitz, Webster, Gebhardt, & Terra, 2000; Gunasekara, 2005). The intellectual tasks and missions of the university require the appreciation of additional sources of funding, but sometimes also the counter-balancing of measures against strong pressures to commercialize.

The saliency of the university within these processes is a consequence of its need to be reflexive on the labor market and therefore adaptive to the conditions in which it has to organize local niches of knowledge-intensive development. The university provides these environments in turn with new (and sometimes counterintuitive) insights. Unlike the systems in its environment, however, the main mover of this system is the continuous flux of students, i.e., human resources. Young people have to be provided with opportunities for further academic careers, functions in the civil service, and perspectives on industrial and entrepreneurial activity. In other words, the university's mission is to recombine these functions (Brewer, Gates, & Goldman, 2001).

*Normative and cognitive orientations in the formation of social capital*
The social systems of science and the clients of the science shops meet at interfaces. However, the two groups can be expected to use very different repertoires and to provide different meanings to the relevant interfaces. First, the two sides are differently organized, potentially (but not necessarily) in terms of institutions. Second, one expects the horizons in the communication to be developed very differently among the partners. Therefore, the science shop functions both as an institution *and* as a mechanism for translation. The relative emphasis on the institution and on its functions can be expected to vary among science shops. The institutions, however, condition how these functions can be fulfilled.

From the perspective of collaboration, the bridging function of the science shop as an institution can be considered as providing an infrastructure for the further development of the knowledge base within and among the three partners involved. These three partners are the clients on the demand side, the researchers and students on the supply side, and the mediators at the level of the science shop. One cannot expect a one-to-one correspondence between the types of knowledge developed and these institutional roles. The parties bring to the interface different elements relevant to the knowledge production process, and they expect different elements in return.

For example, the clients may be able to offer access to domains that are not easily open to questioning without their help. The staff and students at the university are experts on the side of the methodologies and techniques to be used. The theoretical perspectives set constraints which can be expected to vary among disciplines and specialties. For example, using the "grounded theory" approach of social scientists, one would give priority to the reconstruction of the perspective of the subjects under study. The aim is to improve the communicative competences of the partners qualitatively by providing them with (potentially counterintuitive and therefore emancipating) insights. The same report, however, also evaluates and potentially improves the intervention. The reconstruction may provide the different participants with options for change and thus for solving puzzles. The reflexive formulation of these perspectives can help to enlighten other levels of policy-making, but scholars working from within this theoretical tradition will consider this as an indirect effect.

Issues of health and safety are often of a far more objective nature. When an action group comes to a science shop with a worry about the environment (for example, the contamination of water), this group does not expect first to be counseled by the science-shop mediator about subjective worries. The NGOs in this case may wish to use the results at other levels of policy-making. The main interest is then in participation in decision-making, and not in the research process itself. Thus, the nature of the involvement of the citizenship can be expected to vary with the cognitive nature and the social functions of the research questions involved.

The resulting reports and other outcomes have different functions for the three partners involved; these contents have to be "translated" in terms of their relevance for the wider environments. These communication processes involve asymmetries in the understanding that provide sources of potential conflict. The normative integration on each side has to be balanced by rationalized differentiations so that these imminent conflicts can be handled. If the normative integration were to fail on either side, the exchange would become risky and the partners would tend to withdraw from the collaboration. If the rationalized differentiation fails, however, distrust can be expected to emerge because cognitive expectations (as different from normative ones) are damaged. The results of scientific investigations easily lead to counter-intuitive conclusions that need to be extensively explained. In the Berlin case, for example, some results could be used as counter-evidence about the city planning in policy decisions and courtrooms.

The two mechanisms (normative and cognitive expectations) operate at different levels and can disturb a project for very different reasons and with an interaction effect. The failure of normative expectations in the communication functions differently from the failure of cognitive expectations. The latter can be expected to generate a breakdown of trust in the relation, while the former generates anxiety and therefore failure to communicate (Luhmann, 1993). A science shop has to operate and to succeed at both levels. Analytically, this leads to a drive towards professionalization.

The most extreme case of professionalization and institutional independence is illustrated by the report about the science shop in Bonn. This science shop not only functions as an independent association, but has also internalized substantive expertise about the mediating role of a science shop and the factors relevant to it (like funding). The shop publishes independently, and operates outside academia as a consultancy. Most of the science shops, however, are strongly related to universities and do not publish independently because publication is precisely the scholarly task of the scientists (and students) involved. Externally, the main function of the science shop is then to alert the press and to stimulate the publication of brochures. Brochures can be used as pragmatic versions of the (scientific) reports by the client groups.

## 6. Conclusions

The European Commission has now funded a series of three projects. INTERACTS (2001-2003) followed upon SciPas (1999-2001) that focused on making an inventory of ongoing activities, and in 2003 the network ISSNet was also created. The main purpose of INTERACTS was to analyze the practices of science shops substantively by using in-depth case studies. The results of this project enabled us to understand how the transfer of science and technology works at the grassroots level.

The case studies show that despite the variation in terms of nations, disciplines, institutional settings, etc., the science shops have developed a common practice of mediating between citizen groups and the public sphere. However, these practices have evolved in local niches, and have therefore been reflected within these environments (e.g., in brochures in the national language). The communication of science-shop practices beyond these contexts adds another layer to the practical collaboration itself. The comparisons among the case studies have allowed us to distinguish in terms of institutional integration between the higher-education function and the research function of the university. This distinction is perhaps more important than the focus on national differences.

As could be expected, the disciplines vary in terms of how easily they can be accessed at the different levels. In all disciplines, however, the M.A. thesis seems a natural point of access for science-shop questions. Below this level, undergraduates may find projects motivating for engaging in scientific specialization, as in the case of the Danish biology students who involved themselves and their supervisor (!) in the environmental problems of a village. Above this level, some specialties seem more receptive than others for a relation with the local environment. At the social science end, we found the evaluation of social welfare as a point of crystallization, while at the natural science end environmental issues are the prime driver. However, these two extremes only make clear that a host of possibilities exist in between. Thus, a social contract of science and society can no longer be constituted across the board, but should be considered as a variety of communication channels which can be improved. The science-shop model adds the action component to these communication diffusion mechanisms, and thereby potentially generates new stakeholders in the understanding of specific and specialized sources of knowledge.

*Recommendations: communication of the collaborations*

One important recommendation we can provide is in relation to how the science shops can work to ensure the visibility of their project reports. Science shops are sometimes insufficiently aware that the reports have only the status of gray literature within academia; they are not considered as results that can be submitted to scientific journals. Within the scientific production process of scholarly publications, the reports therefore tend to disappear. However, the elaboration into scientific publications and materials for higher education always requires additional reflection. In the case of science-shop projects, this reflection is provided within the scientific institution, but the reflexive communication is then no longer attributed to the science shop. It contains another selection mechanism generated by using the codes of the research system and/or the higher-education system.

Given their increasingly important role in social capital formation, the science shop reports and corresponding M.A. theses deserve to be archived. Archiving and publication on the Internet should further be developed in current science-shop practices. In the science shop case studies, the reports are suspiciously absent. In other circumstances, the materials are linked to assistants and professors who may have disappeared from the institutional presentation of materials on the Internet, perhaps because they left the institution. Because a number of the researchers who acted as supervisors were not on a tenure track, the change of position and therefore visibility should have been anticipated by the local science shops.

In the past attempts have been made to organize the documentation of the science shops centrally. For example, the SciPas project contained an ambitious vision of the creation of a central database by the Loka institute. This database was fully programmed and brought on-line,[4] but the different science shops never utilized it by entering their reports. With hindsight, this project may have been too ambitious and too centralized. The science shops fulfill functions for the respective universities in their local contexts, and it is therefore often important that the reports are profiled as outputs of the respective universities. The local pages can be mirrored (translated, and perhaps classified) at the supra-national level of a central database, but only the mirroring and the further elaboration of search facilities on that basis is then a task for this project. As long as the lower level is not firmly in place, however, the higher-level aggregation can be expected to remain incomplete.

The science shops should be encouraged to make reports available (as some of them are) as files on the Internet. This could be standard practice and one of the criteria for further funding. It would provide researchers, students, and clients with points of reference in their practices, and publication on the Internet can be expected to provide more access and recognition from various sides (Lawrence, 2001). The availability of reports and active updating provide opportunities to claim credit for an innovation at a later stage, even if the effects of the new insights are somewhat disappointing in the short term. For example, if in a later stage (e.g., after the next elections) the municipality of Frederikssund should decide to clean the stinking pools in their village, the reconstruction would become partly attributable to the students who took this initiative. It would be impossible to ignore the link if the reports were properly archived.

What the reports mean on either side of the interface can again be expected to be different. We found several cases where students who wrote a Master's thesis in the framework of the science shop were among the best and were provided with career opportunities at their respective universities. We found also cases where the students left academia, but were included as co-authors in a later publication by the supervisor. The winning of prizes and receiving of awards by both students and scholars involved in science shop work is remarkable. In the Berlin cases this mechanism of recognition and appreciation seems well developed.

The science shops can learn from these mechanisms. One can provide the authors of reports with serious certificates (e.g., from the university) on a yearly basis. A number of "best practices" can also be defined as: "best student paper," "best report," "best advice," etc. A committee could staff the jury where clients, administrators, and scholars (university staff and/or external referees) meet to discuss the results of the past year with the purpose both to provide recognition for the students and scholars involved and to make recommendations for improvements in the quality of the collaboration and the transmission.

---

[4] This database is available at http://www.livingknowledge.org.

      One could further explore the tasks of the science shops in combining normative concerns with analytical perspectives, and the inherent tensions in this type of work could also be made visible. It seems obvious that wherever these mechanisms are successful in solving the puzzles involved, they can be expected to remain fragile. From the perspective of the institutions, the science shops operate at interfaces that are not continuously needed. However, these interfaces may be crucial for the development of a knowledge-based society from a system's perspective. The translation of clients' concerns and demands into the system and the feedback from research and higher education strengthen the social integration of universities and thus provide legitimation for the academic function. This collaboration deeply involves public audiences because their own substantive demands are taken seriously. Academic freedom can thus be appreciated more fully as a societal resource.